\documentclass[%
 reprint,
superscriptaddress,
showpacs,
 amsmath,amssymb,
 aps,
 pra,
floatfix, longbibliography ]{revtex4-1}
\usepackage{graphicx}
\usepackage{dcolumn}
\usepackage{bm}
\usepackage{soul,color}
\usepackage[colorlinks,linkcolor=blue,anchorcolor=blue,citecolor=blue,urlcolor=blue]{hyperref}

\begin{document}

\title{{\hfill\small\rm Phys. Rev. Applied {\bf 13} (2020)}\\%
       {\hfill\small\rm 2020 Collection on {\em 2D Materials and Devices}} \\%
       Periodically Gated Bilayer Graphene as an Electronic Metamaterial
}

\author{Xianqing Lin}
\affiliation{Physics and Astronomy Department,
             Michigan State University,
             East Lansing, Michigan 48824, USA}
\affiliation{College of Science,
             Zhejiang University of Technology,
             Hangzhou 310023, China}

\author{David Tom\'anek}
\email{tomanek@msu.edu} %
\affiliation{Physics and Astronomy Department,
             Michigan State University,
             East Lansing, Michigan 48824, USA}

\date{\today}

\begin{abstract}
We study ballistic transport in periodically gated bilayer
graphene as a candidate for a 2D electronic metamaterial. Our
calculations use the equilibrium Green function formalism and take
into account quantum corrections to charge density changes induced
by a periodically modulated top gate voltage. Our results reveal
an intriguing interference-like pattern, similar to that of a
Fabry-P\'{e}rot interferometer, in the resistance map as a
function of the voltage $V_{BG}$ applied to the extended bottom
gate and $V_{TG}$ applied to the periodic top gate.
\end{abstract}


\maketitle


\section{Introduction}

Photonic metamaterials are artificial structures used to control
propagation of light waves~\cite{Veselago68}. Their
frequency-dependent electromagnetic response in terms of
transmission, reflection and refraction can be tailored using
designer periodic arrays of structural elements spaced closer
than the wavelength of light~\cite{%
{Kosaka1998},%
{Smith2000},%
{Shelby2001},%
{Pendry2006},%
{Stockman04},%
{Yao2008},%
{Ziegler07},%
{valentine2008three},%
{valentine2009optical},%
{Ni2015},%
{Xiang2014},%
{Lv2016},%
{Shi2018}}. %
Same as a photonic metamaterial is capable of manipulating a
coherent electromagnetic wave~\cite{Veselago68}, so should an
electronic metamaterial be able to manipulate a coherent wave of
electrons~\cite{DRAGOMAN1999,Dragoman2007}. Same as propagation of
light can be controlled by periodically modulating the index of
refraction and speed of light $c$ in a three-dimensional (3D)
crystal~\cite{Yao2008,valentine2008three}, so can the propagation
of electrons be controlled by modulating the electrostatic
potential and Fermi velocity $v_F$ in a two-dimensional (2D)
graphene
bilayer~\cite{Young2011,Varlet2014,du2017tuning,kraft2018tailoring}.
Same qualitative behavior should be expected of coherent waves of
electrons and photons with the main difference that the
electrostatic potential is much easier to modulate than the index
of refraction~\cite{Young2011}. Then, phenomena including
scattering, interference, diffraction of light and uncommon
behavior of photons in an optical
metamaterial~\cite{%
{Kosaka1998},%
{Smith2000},%
{Shelby2001},%
{Pendry2006},%
{Yao2008},%
{valentine2008three},%
{valentine2009optical},%
{Ni2015}} should occur on a wider and more flexible range when
manipulating electrons in an electronic metamaterial. In
particular, a periodically gated 2D semiconductor may display the
same transmission behavior for electrons~\cite{esaki73} as a
distributed Bragg reflector (DBR) does for photons~\cite{%
{ReinhartDBR75},{TsangDBR76}}.

To explore the possibility of constructing a 2D electronic
metamaterial, we study theoretically the propagation of electrons
in periodically gated bilayer graphene. Our calculations use the
equilibrium Green function formalism to describe ballistic
transport in bilayer graphene (BLG) and consider quantum
corrections to charge density changes induced by a periodic
modulation of the top gate voltage. Our results reveal an
intriguing interference-like pattern, similar to that of a
Fabry-P\'{e}rot interferometer, in the resistance map as a
function of the voltage $V_{BG}$ applied to the extended bottom
gate and $V_{TG}$ applied to the periodic top gate.

\begin{figure}[t]
\includegraphics[width=1.0\columnwidth]{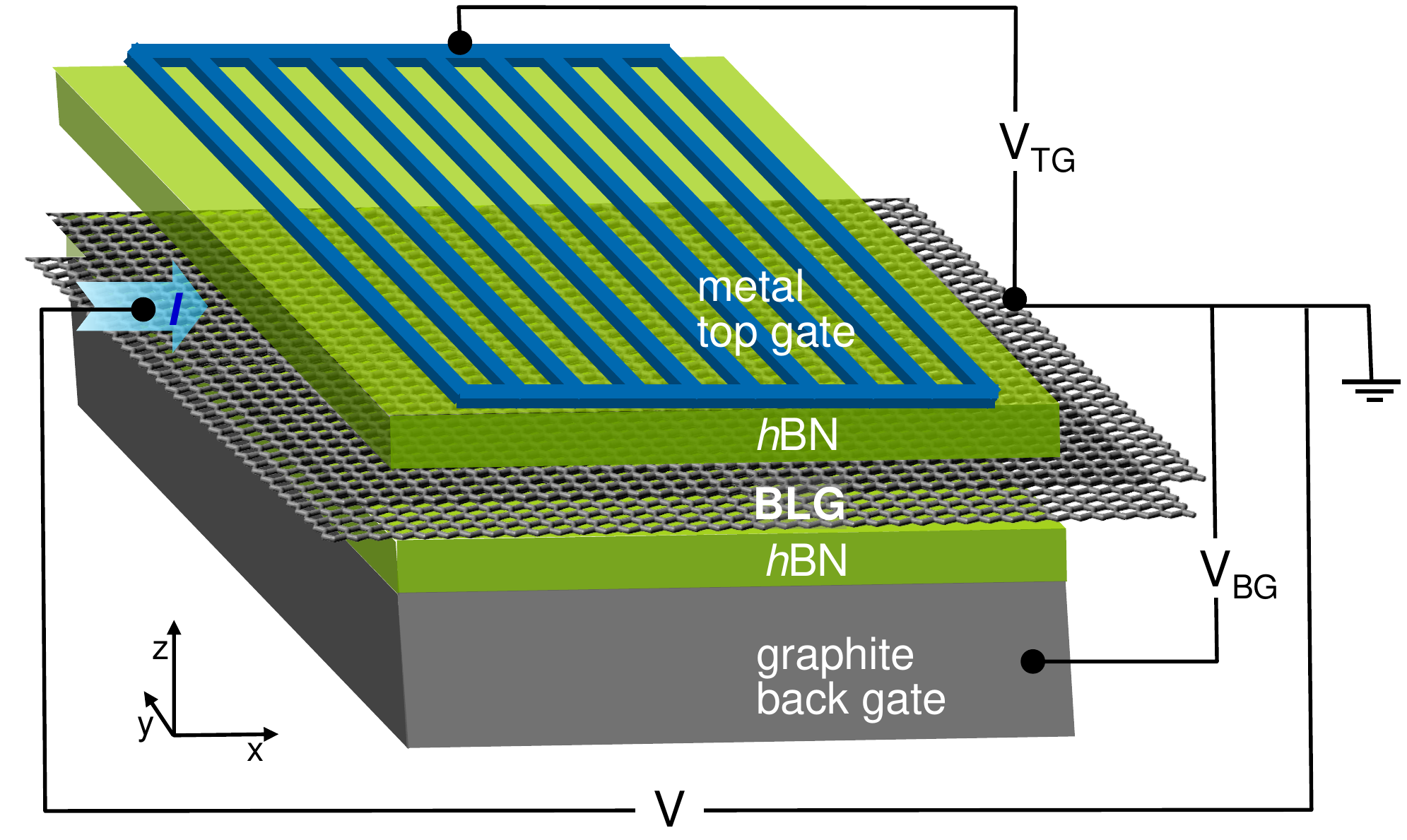}
\caption{%
Schematic view of of a periodically gated bilayer graphene (BLG)
device with a top gate formed by a 2D nanowire array and a
bottom gate extending across the entire device. %
\label{fig1}}
\end{figure}
\begin{figure}[t]
\includegraphics[width=1.0\columnwidth]{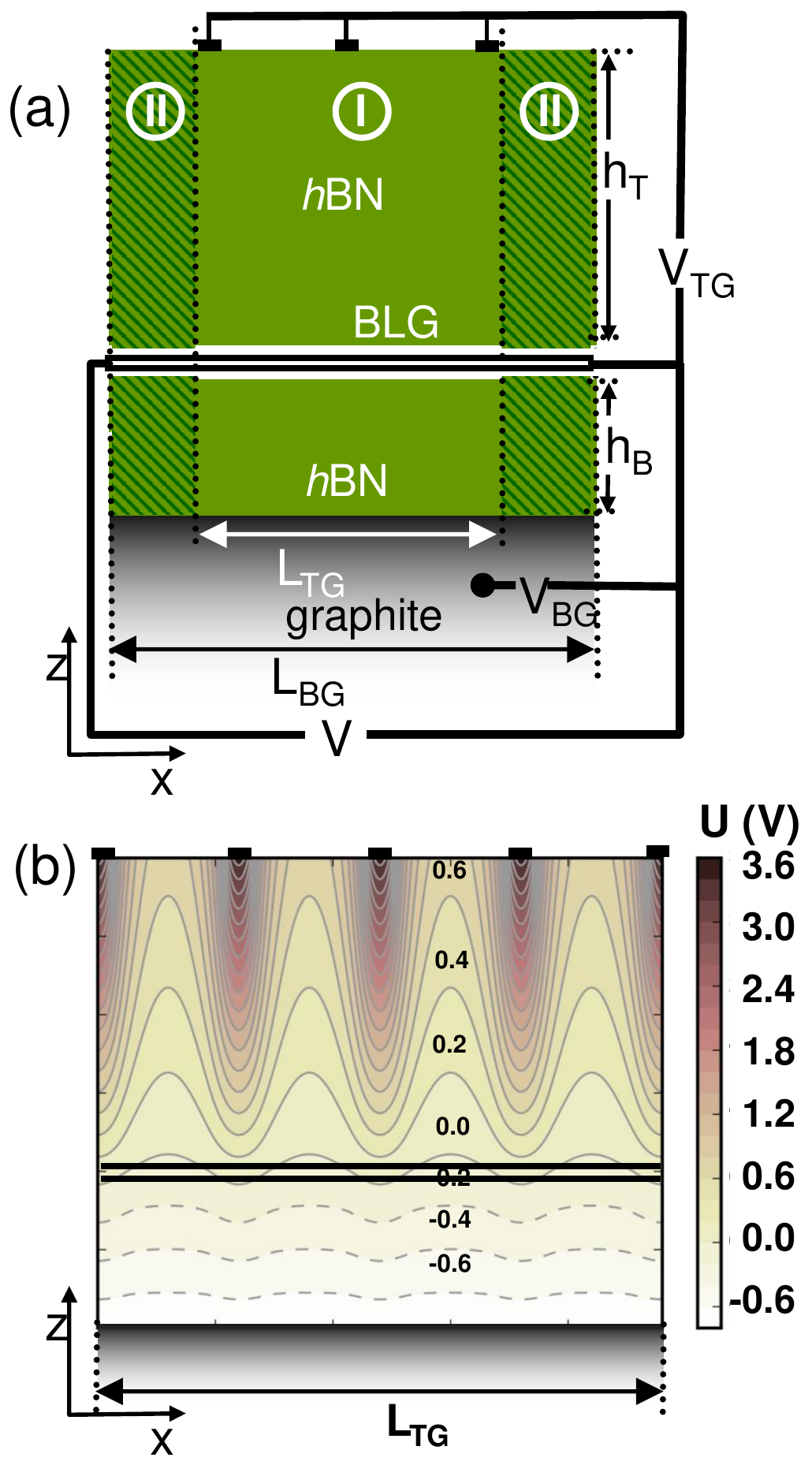}
\caption{Relevant regions and electrostatic potential in a
periodically gated BLG device. %
(a) Schematic side view of the device with BLG in the channel. BLG
is at ground potential and separated from the top gate by an $h_T$
thick $h$BN sheet and from the bottom gate by an $h_B$ thick
protective $h$BN sheet. We distinguish the central region (I)
in-between the top and the bottom gate from region (II) outside
the top gate, but above the bottom gate. %
(b) Electrostatic potential $U(x,z)$ in the central region (I) of
the device calculated for $V_{BG}=-0.76$~V and $V_{TG}=3.48$~V.
\label{fig2}}
\end{figure}

Due to its atomic-scale perfection and unique electronic
structure, monolayer graphene (MLG) has emerged as an ideal 2D
material to study charge transport~\cite{Novoselov04}. Much
attention has been paid to ballistic transport of electrons and
suppression of backscattering by Klein tunneling in MLG, including
the effect of $p-n$ junctions, local and periodic gating,
interaction with the substrate and presence of magnetic
field~\cite{%
{katsnelson2006chiral},%
{Shytov2008},%
{Beenakker2008},%
{young2009quantum},%
{Ramezani2010},%
{Dubey2013},%
{Liu13},%
{yankowitz2012em},%
{ponoclon13},%
{Chen2016},%
{DrienovskyPRB17},%
{DrienovskyArxiv17},%
{drienovsky2018}%
}. %
The band structure of MLG at the six Fermi points in the Brillouin
zone is characterized by Dirac cones, formally describing massless
particles with constant $v_F$ independent of doping.

Our study is devoted to periodically gated BLG which, same as MLG,
is a semimetal. Unlike in MLG, scattering is not suppressed by
Klein tunneling due to the lowered symmetry of BLG. In absence of
Klein tunneling, the resistance of BLG can be tuned to be very
high. The band structure of BLG is qualitatively different from
MLG, as it is characterized by parabolas and not Dirac cones near
$E_F$. Consequently, $v_F$ and thus the wavelength of electrons
can be modulated by local doping caused by changing the
electrostatic potential. Thus, the BLG system appears to be a
better candidate for electron optics than MLG, which we will also
discuss for the sake of reference. The possibility of constructing
the electronic counterpart of an optical DBR has not been explored
so far.

\section{Structure and Functionality of Periodically Gated BLG}

The schematic of a recently fabricated
device~\cite{XiangZhang-expt}, consisting of periodically gated
BLG sandwiched in-between inert $h$BN layers, is presented in
Fig.~\ref{fig1}. The BLG channel is contacted by metal leads at
the source and the drain ends and is separated by an $h_B=10$~nm
thick $h$BN layer from the bottom electrode and by an $h_T=20$~nm
thick $h$BN layer from the top electrode. The top electrode
consists of a periodic array of parallel, $W=25$~nm wide wires,
separated by $L=120$~nm. The bottom gate voltage $V_{BG}$
regulates the doping level of the channel, whereas the top gate
voltage $V_{TG}$ modulates the electrostatic potential along the
channel. The device performance is characterized by the resistance
$R$ between source and drain.

To provide an adequate description of the gated BLG device under
operating conditions, we distinguish its components and their
function in the schematic cross-section provided in
Fig.~\ref{fig2}(a). Only the central region of the device, labeled
(I), lies between the non-uniform top gate (TG) of length
$L_{TG}$, formed of a metal wire array, and the bottom gate (BG)
of length $L_{BG}>L_{TG}$, formed of a graphite slab. This is the
region of interest for electron optics to be discussed below.

Even though region (II), which lies in-between region (I) and the
contacts, may be of lesser interest, it still needs to be
addressed in the transport study. This region is above the BG and
thus affected by $V_{BG}$, but outside the range of the TG and
thus unaffected by $V_{TG}$. Key to the interpretation of the
resistance in region (II) is the interface between BLG and $h$BN
layers above and below the channel. There is only negligible
electronic interaction between graphene layers and the surrounding
$h$BN due to its $5.97$~eV wide band gap~\cite{Watanabe2004}. Even
if the BLG were perfectly aligned with $h$BN, the 1.8\% lattice
mismatch would give rise to a Moir\'{e}
superlattice~\cite{yankowitz2012em,ponoclon13}. Minor lattice
relaxation in the graphene layer caused by their interaction with
$h$BN would then modulate periodically the potential in the
graphitic channel, giving rise to second-generation Dirac
points~\cite{%
{yankowitz2012em},%
{ponoclon13},%
{dean2013hofstadter},%
{jung2015origin},%
{Wallbank2015}%
}. %
In perfectly aligned BLG/$h$BN superlattices, we expect the
electronic density of states (DOS) to vanish at $E_F$ as a
consequence of first-generation Dirac points at the charge
neutrality level and at ${\Delta}E<<1$~eV below and above $E_F$ as
a consequence of newly formed second-generation Dirac points. For
$V_{BG}=0$, $E_F$ is located at first-generation Dirac points,
resulting in high resistance that is independent of $V_{TG}$ and
represented by a line in the $R(V_{TG},V_{BG})$ resistance map.
Applying a bottom gate voltage $V_{BG}$ induces a nonzero charge
density $\sigma_{BLG}={\epsilon}V_{BG}/h_B$ in the channel, where
${\epsilon}$ is the dielectric constant and $h_B$ is the thickness
of the lower $h$BN layer, as defined in Fig.~\ref{fig2}(a). We
find that the charge density needed to reach the secondary Dirac
points may be induced by $V_{BG}=-1.5$~V when using
${\epsilon}{\approx}7.0\epsilon_0$~\cite{epsilon-hBN66} and
$h_B=10$~nm in the BLG device. The large resistance at this value
of $V_{BG}$ is again independent of $V_{TG}$, giving rise to a
second parallel line in the $R(V_{TG},V_{BG})$ resistance. For
voltages other than $V_{BG}=0$~V and $-1.5$~V, the resistance map
reflects only the behavior in region (I).

\section{Results}

\subsection{Transport in Periodically Gated BLG at $T=0$}

To determine the resistance pattern associated with the central
region (I) of interest, we first calculate the electronic
structure of BLG and the electrostatic potential $U$ within the
plane of the channel as a function of $V_{TG}$ and $V_{BG}$. For a
given $V_{BG}-V_{TG}$ combination, the propagation of ballistic
electrons and the net resistance of the gated BLG device is
evaluated using the equilibrium Green function formalism.

As indicated in Figs.~\ref{fig1} and \ref{fig2}(a), we denote the
transport direction $x$ and the direction of the TG wires by $y$.
The width $W_{TG}$ and length $L_{TG}$ of the periodically gated
region is much larger than any other dimensions in the device and
may be considered infinite. Due to this large size, atomistic
calculation of the entire structure is out of the question and
would only complicate the interpretation of transport results in
periodically gated BLG. In the cryogenic regime with a very small
applied source-bias voltage, transport in the BLG channel can be
considered to be ballistic and attributed to propagation of
low-energy charge carriers in a periodically modulated potential
$U(x)$.

The low-energy Hamiltonian of a free-standing, ungated BLG can be
written as~\cite{McCann2013}
 \begin{equation}
 H_{BLG}(k_x,k_y) = \left( %
 {\begin{array}{*{20}{c}} %
 0 & v_F{{\bf{p}}}^\dag & 0 & 0\\
 v_F{{\bf{p}}} & 0 & \gamma_1 & 0\\
 0 & \gamma_1 & 0 &v_F{{\bf{p}}}^\dag \\
 0 & 0 & v_F{{\bf{p}}} & 0
 \end{array}} %
 \right) . %
 \label{eq1}
 \end{equation}
Here we use ${{\bf{p}}}=\hbar(k_x+ik_y)$ with $(k_x,k_y)$ to
describe the carrier momentum with respect to the Fermi momentum
at the Fermi point in the corner of the hexagonal Brillouin zone.
The tight-binding parameters describing these systems
are~\cite{DT029} the intra-layer nearest neighbor $pp{\pi}$
hopping integral $\gamma_0=-2.66$~eV and the inter-layer nearest
neighbor $pp\sigma$ hopping integral $\gamma_1=0.27$~eV. This
yields $\hbar{v_F}={3/2}(-\gamma_0)d$, where $d=1.42$~{\AA} is the
intra-layer nearest neighbor distance. Only the diagonal matrix
elements will be affected by the modulation of the potential in
the field of the periodic top gate, since the top gate period $L$
is much larger than the interatomic spacing.

The two low-energy bands of $H_{BLG}$ are
\begin{equation}
E_{\pm}(k) = %
\pm\frac{1}{2}(-\gamma_1 + \sqrt{4\hbar^2v_F^2k^2+\gamma_1^2}) , %
\label{eq2}
\end{equation}
where $k=\sqrt{k_x^2+k_y^2}$ is close to the Fermi momentum $k_F$.
$E_{+}(k)$ describes the dispersion in the conduction band and
$E_{-}(k)$ that in the valence band.

In BLG gated by a periodic top and a uniform bottom gate, the net
electron number density $n(x)$ varies periodically along the
transport direction and is constant in the $y$ direction. In BLG
with isotropic band dispersion at $E_F$, we find
\begin{equation}
n(x) = sign(n)\ {k_{F}^2}(x)/{\pi} \,, %
\label{eq3}
\end{equation}
where $k_{F}(x)$ is the Fermi wavevector at position $x$. There is
particle-hole symmetry with positive $n$ for electron and negative
$n$ for hole doping.

The dependence of the charge density $n$ and the Fermi momentum
$k_F$ on $x$ is in response to the periodic electrostatic
potential $U(x)$ in the plane of the BLG. With the contact lead at
the drain end at ground potential, which sets $E_F=0$ within the
BLG, this potential is given by
\begin{equation}
(-e)U(x) = -E_{\eta}\left(k_{F}(x)\right)\,, %
\label{eq4}
\end{equation}
where $e$ is the absolute value of the electron charge. The
subscript ${\eta}$ in the expression for $E_{\eta}$ in
Eq.~(\ref{eq2}) is either $+$ in case of electron doping or $-$ in
case of hole doping. The sign of $U(x)$ is the same as that of
$E_{\eta}(k_{F}(x))$ and $n(x)$.

In principle, $U(x)$ could be obtained for any gate geometry by
solving the Poisson equation~\cite{Liu13}. To avoid this
calculation for every combination of $V_{BG}$ and $V_{TG}$, we use
an alternate approach. We note that in BLG exposed to the periodic
electrostatic potential $U(x)$ caused by the TG voltage $V_{TG}$
and the BG voltage $V_{BG}$, $n(x)$ can be expressed by
\begin{equation}
n(x) = -\frac{1}{e} %
     \bigl(%
     C_{T}(x)\bigl[U(x)-V_{TG}\bigr]
    +C_{B}(x)\bigl[U(x)-V_{BG}\bigr]%
     \bigr)\,. %
\label{eq5}
\end{equation}
Here, the doping charge density $n(x)$ has been related to changes
in the potential by the position-dependent partial
capacitances~\cite{Liu13} $C_{T}(x)$ of the top gate and $C_{B}$
of the bottom gate. The above expression can be rewritten as
\begin{eqnarray}
n(x) &=& \frac{1}{e}\left( C_{T}(x)V_{TG}+C_{B}V_{BG} \right)
       \nonumber \\
     &\ & +\frac{1}{e}\left( C_{T}(x)V_0 + C_{B}V_0 \right)
           \frac{U(x)}{V_0}
       \nonumber \\
      &=& n_c(V_{BG},V_{TG},x) %
         +n_c(V_0,V_0,x)\frac{U(x)}{V_0} \,, %
\label{eq6}
\end{eqnarray}
which defines a new quantity, namely the classical net electron
number density $n_c$. This quantity depends on the position $x$
within the BLG, considered to be a classical metal, the gate
geometry and the gate voltages $V_{TG}$ and $V_{BG}$. $n_c$ is
nominally defined by
$n_c(V_{BG},V_{TG},x)=(1/e)[C_{T}(x)V_{TG}+C_{B}V_{BG}]$ and can
be calculated in the BLG plane using classical electrostatics for
the specific gate geometry. For a given charge density
$\sigma_{T}$ distributed uniformly across the top gate wires,
which are separated by a dielectric of thickness $h_T$ and
dielectric constant $\epsilon$ from the grounded BLG, we can
numerically determine $n_c(x)$ and the electric field in the
entire region using the image-charge technique, which also
guarantees a constant zero potential in the BLG layer. Integrating
the electric field between the TG and the BLG yields the
corresponding value of $V_{TG}$, and the same approach can be used
for the bottom gate. We note that $V_{TG}$ is proportional to
$\sigma_{T}$ and $V_{BG}$ is proportional to $\sigma_{B}$,
providing quantitative values for $C_{T}(x)$ and $C_{B}$. $V_0$ is
a nominal voltage value taken to be $V_0=1$~V.

\begin{figure}[t]
\includegraphics[width=1.0\columnwidth]{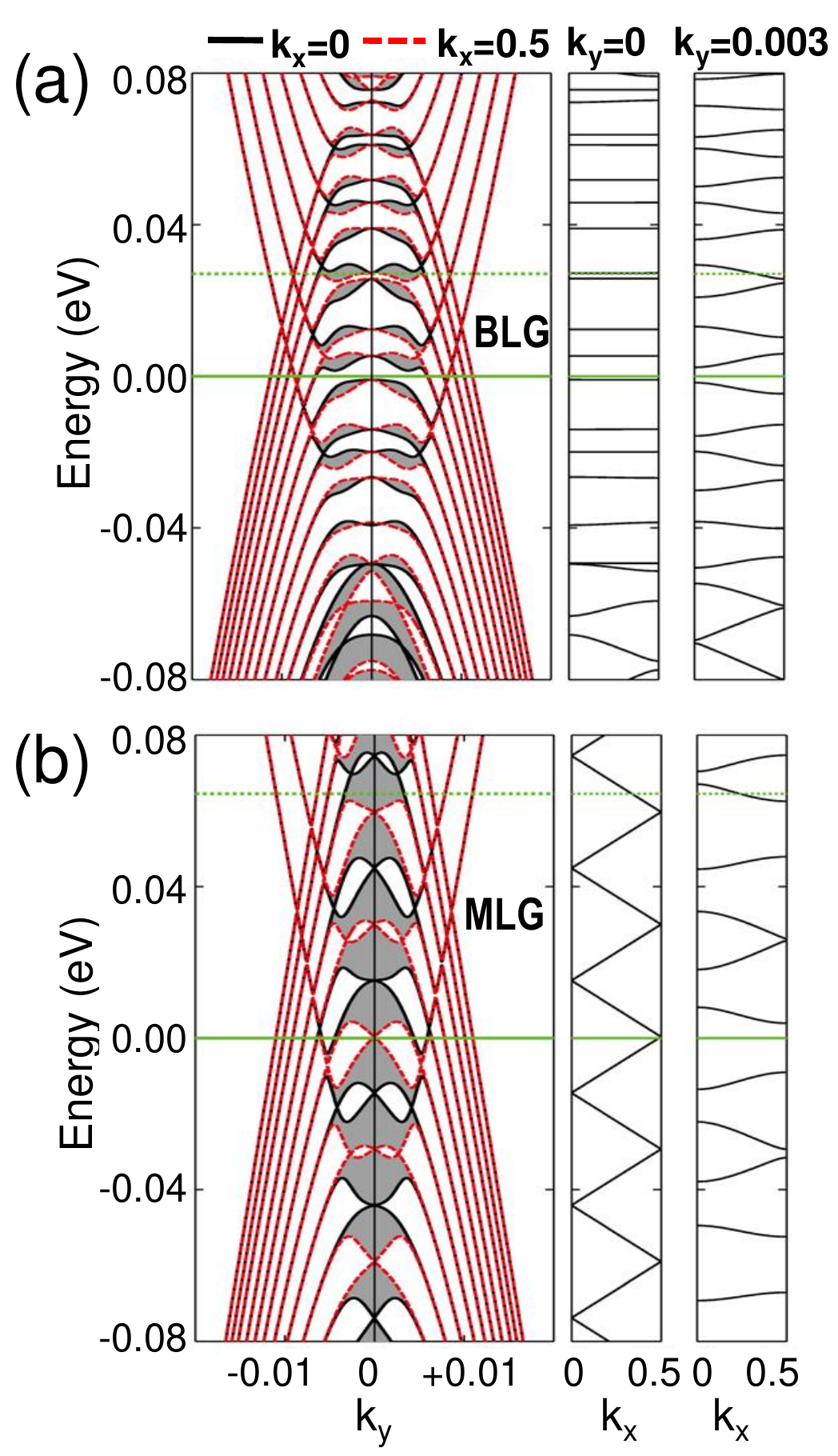}
\caption{Electronic band structure of gated %
(a) BLG and %
(b) MLG. %
Presented $E(k)$ results are obtained for %
BLG with $V_{TG}=4.76$~V and $V_{BG}=-1.34$~V in panel (a) and %
MLG with $V_{TG}=4.42$~V and $V_{BG}=-1.36$~V in panel (b). %
$k_x$ is given in units of $2{\pi}/L$ and %
$k_y$ in units of $2{\pi}/(\sqrt{3}d)$, where
$d=1.42$~{\AA} is the interatomic distance in graphene. %
In the left panels of (a) and (b), %
the black solid lines denote bands with $k_x=0$ and %
the red dashed lines denote bands with $k_x=0.5$ %
at the Brillouin zone boundary. %
The green solid lines denote the $E=E_F=0$ energy level and the %
green dashed lines denote the gate-dependent charge-neutrality %
level in each system. %
The two right panels in (a) and (b) show band dispersion along
$k_x$ for two values of $k_y$. %
\label{fig3}}
\end{figure}

Substituting Eqs.~(\ref{eq4}) and (\ref{eq6}) into
Eq.~(\ref{eq3}), we obtain an equation for $k_F$ as a function of
$x$
\begin{eqnarray}
n_c(V_{BG},V_{TG},x) + n_c(V_0,V_0,x)
 \frac{E_{\eta}(k_{F}(x))}{e~V_0}
 \nonumber \\
= sign(n_c) \frac{k_{F}^2(x)}{\pi}, %
\label{eq7}
\end{eqnarray}
where $\eta = sign\left(n_c\left(V_{BG},V_{TG},x\right)\right)$.

We note that considering the grounded BLG channel as a classical
metal, nonzero $V_{TG}$ and $V_{BG}$ can induce periodic variation
in the classical density $n_c(x)=n_c(V_{BG},V_{TG},x)$ while
keeping the electrostatic potential $U(x,z=0)$ constant within the
BLG. The fact that BLG is not a classical metal, but a semi-metal
with a vanishing DOS at $E_F$, necessitates further consideration.
Unlike in a classical metal with a large DOS at $E_F$, periodic
variations of $n_c(x)$ in BLG with a small DOS at $E_F$ will cause
a nominal periodic modulation of $E_F$. To keep $E_F$ constant,
the classical carrier density $n_c$ within the BLG will be
modified by what we call a quantum correction $n(x)-n_c(x)$. In
this better description, the periodic electrostatic potential
$U(x,z=0)$ in the semimetallic BLG is no longer constant and will
play an important role. Then, also $U(x,z)$ for a given $z$
between the BG and the BLG will no longer be constant. In the
region between the TG and BLG, quantum corrections dampen the
oscillations in $U(x,z)$ at constant $z$. The electrostatic
potential $U(x,z)$ associated with the quantum corrected carrier
density $n(x)$ within the BLG, caused by $V_{TG}=3.48$~V and
$V_{BG}=-0.76$~V, is shown in Fig.~\ref{fig2}(b) for the central
region (I) and $z$ between the TG and the bottom gate.

Being able to determine the electrostatic potential $U(x)$ and the
position-dependent Fermi momentum $k_F(x)$ using Eq.~(\ref{eq7}),
we can express the position-dependent potential energy $\phi(x)$
of low-energy electrons or holes in BLG by
\begin{equation}
\phi(x)=(-e)U(x)=-E_{\eta}\left(k_{F}\left(x\right)\right)\,.
\label{eq8}
\end{equation}

For BLG in the periodic potential energy surface $\phi(x)$, the
system becomes a superlattice with the lattice constant $L$ along
the $x$ direction, with $L=120$~nm for the device shown in
Fig.~\ref{fig1}. The low-energy bands of this superlattice are
given by the eigenvalues of
\begin{equation}
 \tilde{H}_{BLG} = \left( %
 {\begin{array}{*{20}{c}} %
 \phi(x) & v_F{\bf{p}}^\dag & 0 & 0\\
 v_F{\bf{p}} & \phi(x) & \gamma_1 & 0\\
 0 & \gamma_1 & \phi(x)  & v_F{\bf{p}}^\dag \\
 0 & 0 & v_F{\bf{p}} & \phi(x) \,.
 \end{array}} %
 \right) %
\label{eq9}
\end{equation}
Here, the wavevector ${\bf{p}}$ with respect to the Fermi
momentum, defined in Eq.~(\ref{eq1}), has become the operator
${\bf{p}}=(-i\partial/\partial{x}+ik_y)$ due to the $x-$dependence
of the diagonal elements. Since $\phi(x)$ varies very slowly and
thus can be represented by only a small number of Fourier
components, $\tilde{H}_{BLG}$ can be diagonalized using as basis
the eigenfunctions of the free-standing $H_{BLG}(\tilde{k}_x,k_y)$
with the momentum vectors in the superlattice geometry.

The electronic band structure of the gated BLG is presented in
Fig.~\ref{fig3}(a) for representative values $V_{TG}=4.76$~V and
$V_{BG}=-1.34$~V. The data presented in the left panel display
$E(k_x,k_y)$ at two values of $k_x$ in the short Brillouin zone of
the superlattice. The shaded regions in-between the bands indicate
the range of band dispersion and white regions indicate local band
gaps. We note that BLG becomes charge neutral when all bands below
the charge-neutrality level, shown by the green dashed line,
become occupied. The band dispersion along $k_x$, shown for two
$k_y$ values in the two right panels, indicates that bands are
almost flat and separated by gaps near the zero-energy level. We
find that at other values of $V_{TG}$ and $V_{BG}$ the band
structure is qualitatively very similar, but shifts periodically
with respect to $E_F$.

\begin{figure}[t]
\includegraphics[width=1.0\columnwidth]{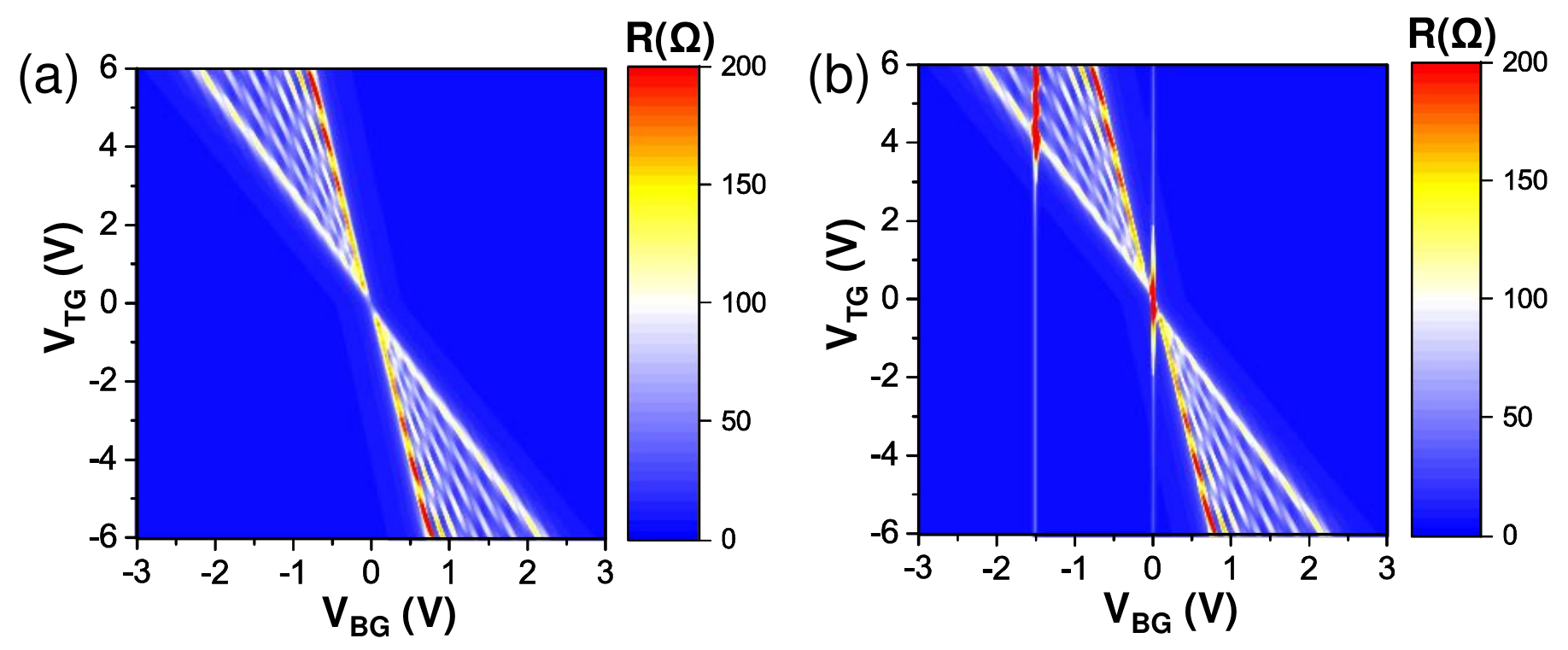}
\caption{%
Calculated resistance map of the BLG device. %
(a) Calculated source-drain resistance $R$ within the central
region (I) as a function of the top gate voltage $V_{TG}$ and the
bottom gate voltage $V_{BG}$. %
(b) Predicted resistance map of the entire device containing both
regions (I) and (II). Two vertical lines at $V_{BG}=0$ and
$V_{BG}{\approx}-1.5$~V associated with first- and
second-generation Dirac points in region (II) are superimposed to
the resistance gap of region (I) in (a). %
\label{fig4} }
\end{figure}

Conductance $G$ is known to be quantized in a system with a finite
cross-section in the ballistic regime~\cite{Datta-book}. To
interpret transport in the device we investigate, we need to
consider its finite width $W_{TG}=11.4~\mu$m. Allowed eigenstates
will then be standing waves normal to the transport direction and
$k_y$ will be quantized. For each $k_y$ value, every band that
crosses the zero-energy level along $k_x$ of the superlattice
provides one conduction channel. Each conduction channel
contributes a conductance quantum
$G_0=2e^2/h=(12.9~k\Omega)^{-1}$. Then, the total conductance $G$
is obtained by counting the number of $k_y$ values associated with
bands dispersing along the $k_x$ direction that cross the
zero-energy level. The total number of conduction channels in the
real device of width $W_{TG}$ becomes
\begin{equation}
M=\sum_{k_y} f(k_y) \,, %
\label{eq10}
\end{equation}
where allowed $k_y$ values are integer multiples of
$2{\pi}/W_{TG}$ and $f(k_y)$ is the average transmission
probability per $k_y$ mode. At $T=0$~K, an allowed state with
given $(k_x,k_y)$ is either occupied or empty. In that case, it
will fully contribute to transmission with probability $f(k_y)=1$
if a band crosses the zero-energy level along $k_x$ for a given
$k_y$ value, or otherwise not contribute at all, so that
$f(k_y)=0$.

Transport calculations for a ballistic device at a non-zero
source-drain voltage $V_{sd}$ are typically performed using the
non-equilibrium Green function formalism~\cite{NEGF2003}. In the
device we consider, which is driven by a source of very small
constant current, $V_{sd}$ is negligibly small. In that case,
transport can be calculated using the equilibrium Green function
that describes the electronic structure of the unperturbed system.

The resistance of the central region (I) is then given by
$R=G^{-1}=(MG_0)^{-1}$. To obtain a smooth map of $R$ at $T=0$~K
as a function of $V_{TG}$ and $V_{BG}$, we have convoluted the
conductance $G(V_{BG},V_{TG})$ with a Gaussian function at each
$V_{BG}$ and obtained
\begin{equation}
\tilde{G}(V_{BG},V_{TG}) %
      = \frac{1}{{\sigma}\sqrt{2{\pi}}} %
        \int G(V_{BG},V) %
        e^{-\frac{(V-V_{TG})^2}{2\sigma^2}} dV \,,
\label{eq11}
\end{equation}
where $2.355~{\sigma}$ is the full width at half maximum of the
Gaussian function. The smooth resistance map
$\tilde{R}(V_{BG},V_{TG}) = 1/\tilde{G}(V_{BG},V_{TG})$ is then
obtained and compared with the experimental results.

Figure~\ref{fig4}(a) shows the calculated smooth resistance map of
the central region (I) and Fig.~\ref{fig4}(b) that of the entire
device with $L=120$~nm and $W=25$~nm.

Electrons are doped into BLG at positive gate voltages and holes
at negative gate voltages. At negative bottom gate voltages
$V_{BG}$ and negative or small positive top gate voltages
$V_{TG}$, BLG is hole doped everywhere and thus shows low
resistance, represented by the uniform dark blue color of the
bottom left region of the resistance map in Fig.~\ref{fig4}(a). At
large positive values of $V_{TG}$ and $V_{BG}$, on the other hand,
BLG is electron doped everywhere and thus also shows low
resistance, as indicated by the same dark blue color of the top
right region in the resistance map. At given $V_{BG}<0$ combined
with moderate $V_{TG}>0$ values, and alternately at given
$V_{BG}>0$ combined with moderate $V_{TG}<0$ values, regions of
hole and electron doping in BLG alternate along the transport
direction $x$. In that case, also the sign of $n(x)$ and $\phi(x)$
alternates along the $x$-direction and electrons, which have been
injected at the zero-energy level $E_F=0$ at the source contact,
have to tunnel through a periodic array of potential barriers.
Then, constructive or destructive interference may cause
significant oscillations in the net resistance as seen in
Fig.~\ref{fig4}(a), similar to a Fabry-P\'{e}rot interferometer.

\begin{figure}[t]
\includegraphics[width=1.0\columnwidth]{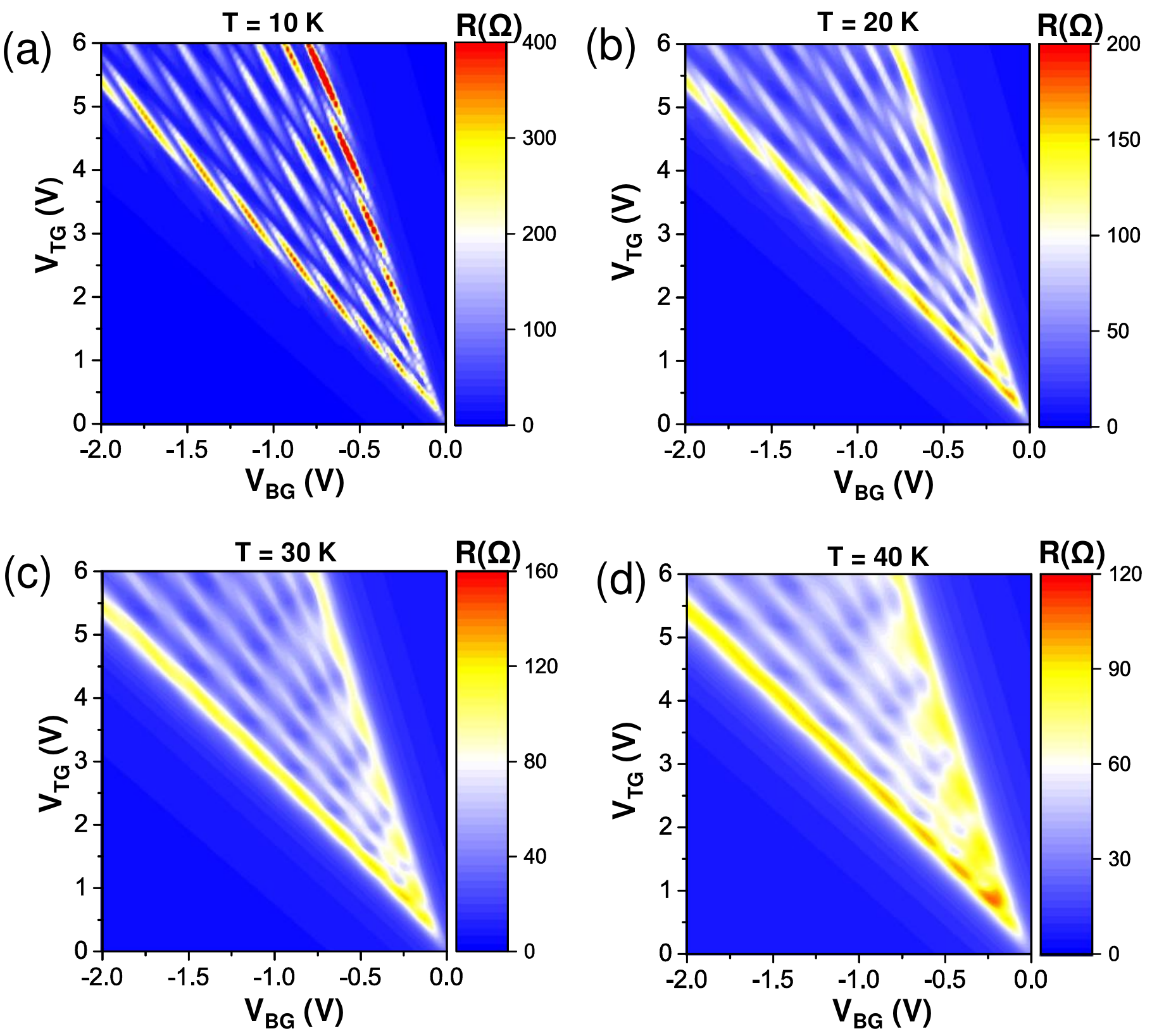}
\caption{Calculated resistance maps of the BLG device at
temperatures
(a) $T=10$~K, %
(b) $T=20$~K, %
(c) $T=30$~K, and %
(d) $T=40$~K. %
\label{fig5}}
\end{figure}

In order to understand the origin of oscillations of
$R(V_{TG},V_{BG})$ in the resistance map, we refer to the
calculated band structure of the BLG superlattice in a constant
potential, shown in  Fig.~\ref{fig3}(a). Results shown in the
middle panel of Fig.~\ref{fig3}(a) indicate that the band
dispersion along $k_x$ near $k_y=0$ of the superlattice is very
small. The white regions in the left panel of Fig.~\ref{fig3}(a)
correspond to band gaps near $k_y=0$, which are not affected by
this small band dispersion along $k_x$. When the zero-energy level
$E_F$ lies in such a local gap, electrons injected at $E_F$ can
not propagate, corresponding to a high resistance. At somewhat
larger $k_y$ values such as $|k_y|=0.003$, the band dispersion
along $k_x$ increases, as seen in the right panel of
Fig.~\ref{fig3}(a). In that case, a momentum $(k_x,k_y)$ may be
found, at which a band crosses $E_F$, thus forming a conductance
channel and reducing the resistance. As seen in the left panel of
Fig.~\ref{fig3}(a), the band dispersion along $k_x$ decreases
again at still larger values of $|k_y|$, thus lowering the
likelihood of transmission and increasing the resistance. As
mentioned earlier, this discussion considered charge transport in
the special case of a constant potential. Changing the gate
voltages changes and modulates the potential along the transport
direction. Gradual changes in the potential move locally the band
structure up or down in energy with respect to $E_F$, thus
changing the number of bands crossing $E_F$ along $x$. A
transmission channel will only then contribute a conductance
quantum if it is open for all values of $x$. The above reasoning
explains the appearance of alternating conductance and resistance
maxima associated with changing gate voltages.

\subsection{Effect of Temperature on Transport in Periodically Gated BLG}

Unlike at $T=0$~K discussed so far, allowed $(k_x,k_y)$ states
near $E_F$ may be partially occupied by the Fermi-Dirac
distribution at $T>0$. Then, the average transmission probability
$f(k_y)$ per $k_y$ mode, introduced in Eq.~(\ref{eq10}), may take
a value in the entire range $0{\le}f{\le}1$ for each band along
$k_x$. Accommodating the band dispersion along $k_x$, we
find~\cite{Datta-book}
\begin{equation}
f(k_y) = \sum_{m} %
         \left(
         \frac{1}{e^{\frac{E^{\rm{min}}_m(k_y)}{{k_B}T}}+1}%
        -\frac{1}{e^{\frac{E^{\rm{max}}_m(k_y)}{{k_B}T}}+1}%
         \right) \,, %
\label{eq12}
\end{equation}
where we have given all energies with respect to $E_F=0$. We have
further noted a near-linear dispersion of the $m$-th band along
$k_x$, ranging from $E^{\rm{min}}_{m}(k_y)$ to
$E^{\rm{max}}_{m}(k_y)$, for a given value of $k_y$. $k_B$ is the
Boltzmann constant.

The effect of temperature on the resistance map, traced back to
the temperature dependence of the channel transmission probability
in Eq.~(\ref{eq12}), is shown in Fig.~\ref{fig5}. Results for
identical gate geometries indicate no net shifts, but just thermal
smearing of $R(V_{TG},V_{BG})$.

\subsection{Effect of Geometry on Transport in Periodically Gated BLG}

\begin{figure}[t]
\includegraphics[width=1.0\columnwidth]{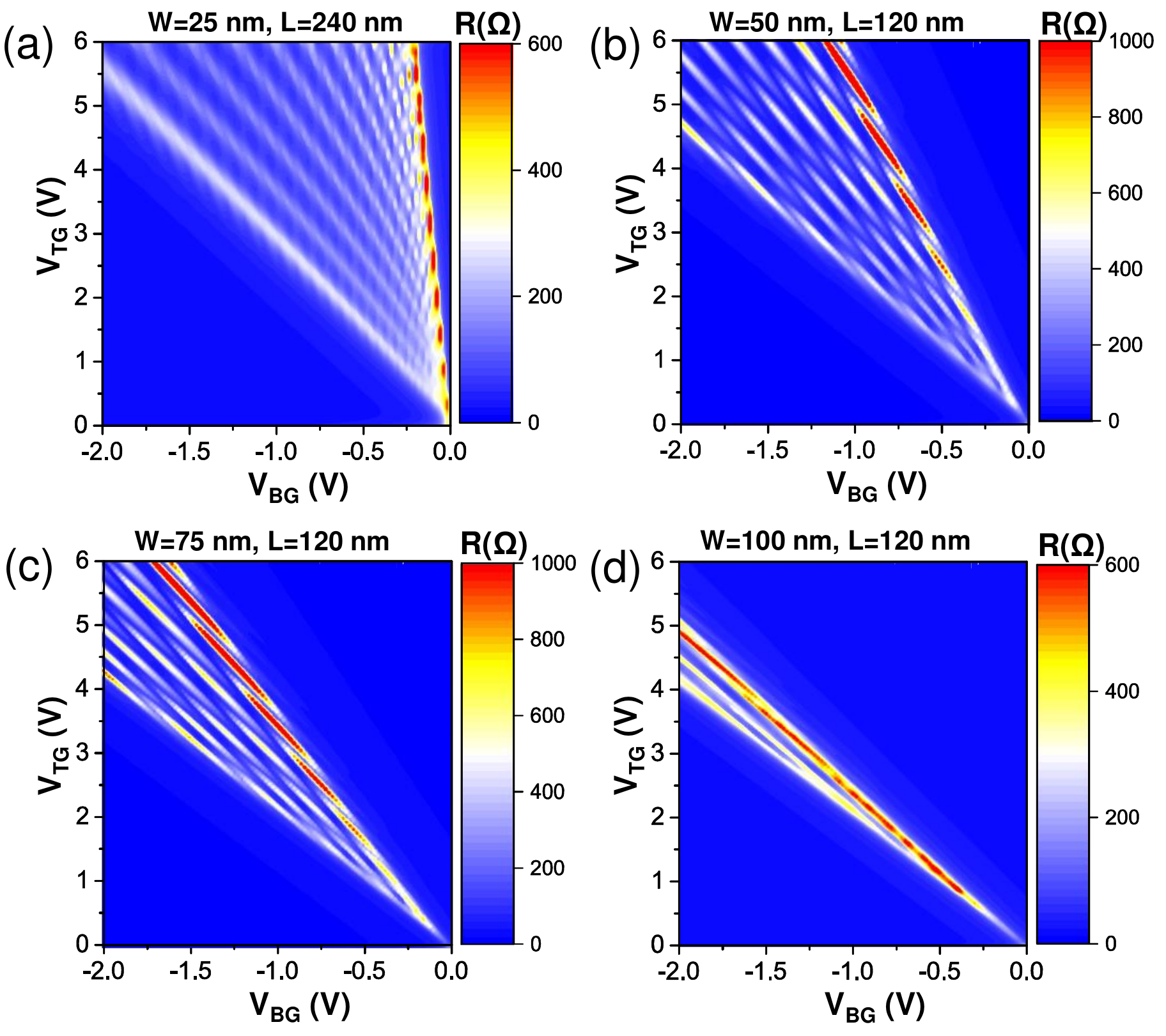}
\caption{%
Calculated resistance maps of BLG devices with different values of
the width $W$ of each wire and the inter-wire distance $L$ within
the periodic top gate. Presented are results for %
(a) $W=25$~nm, $L=240$~nm, %
(b) $W=50$~nm, $L=120$~nm, %
(c) $W=75$~nm, $L=120$~nm, and %
(d) $W=100$~nm, $L=120$~nm. %
Numerical data in the respective panels have been convoluted by
Gaussians with %
(a) ${\sigma}=0.12$~V, %
(b) ${\sigma}=0.06$~V, %
(c) ${\sigma}=0.04$~V, and %
(d) ${\sigma}=0.04$~V. %
\label{fig6}}
\end{figure}

The resistance map $R(V_{TG},V_{BG})$ also depends on the geometry
of the BLG device. To inspect this dependence, we present in
Fig.~\ref{fig6} the calculated resistance map of BLG devices with
different values of the width $W$ of each wire and the inter-wire
distance $L$ within the periodic top gate. As seen in
Fig.~\ref{fig6}(a), high-resistance lines become continuous in
case that $L>>W$. Results in Fig.~\ref{fig6}(b)-\ref{fig6}(d)
indicate that for a fixed $L$, the series of high-resistance lines
tilts and their number decreases with increasing $W$.

\subsection{Comparison with Periodically Gated MLG}

\begin{figure}[t]
\includegraphics[width=1.0\columnwidth]{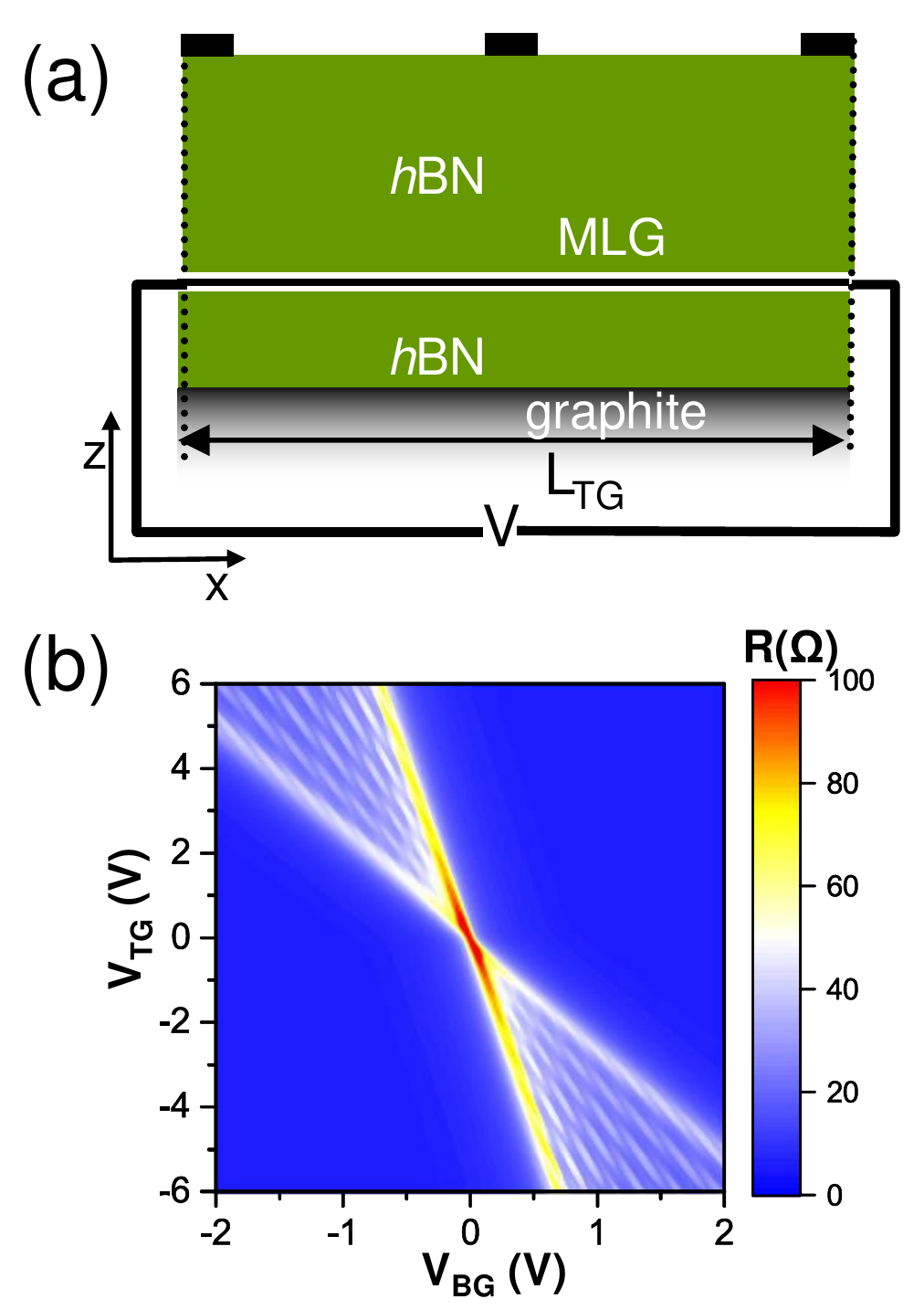}
\caption{%
(a) Schematic cross-section and %
(b) the calculated resistance map of the MLG device. %
$R(V_{TG},V_{BG})$ results have been convoluted by a Gaussian with
full-width at half maximum of $0.05$~V. %
\label{fig7}}
\end{figure}

As a matter of reference, we compare in the following our results
for BLG to MLG in the same device geometry, depicted in
Fig.~\ref{fig7}(a). In consideration of the absence of Klein
tunnelling in BLG due to the interlayer coupling, we have used BLG
rather than monolayer graphene (MLG) here as the channel to
demonstrate such a resistance map, as analyzed below. The only
difference between the MLG and BLG device is the simpler
Hamiltonian, which is given in analogy to Eq.~(\ref{eq1}) by
\begin{equation}
H_{MLG} = \left( %
{\begin{array}%
{*{20}{c}} %
0 & v_F {\bf{p}}^\dag \\
v_F {\bf{p}} & 0
\end{array}} %
\right) %
\label{eq13}
\end{equation}
and which leads to the band structure presented in
Fig.~\ref{fig3}(b).

The MLG-based device we consider is nearly identical to that shown
in Figs.~\ref{fig1} and \ref{fig2}(a), but with BLG replaced by
MLG as the channel. The calculated resistance map of the MLG-based
device, shown in Fig.~\ref{fig7}(b), displays a similar
$R(V_{TG},V_{BG})$ pattern as the BLG device. A notable difference
between the two is a much lower contrast in the MLG than in the
BLG device, with the resistance peak values for MLG being much
lower. In addition, compared with MLG, the resistance peaks for
BLG are wider and higher, and the resistance valleys are also
wider and shallower. We also note that in the resistance map
without convolution, the MLG device has many more resistance
peaks. These peaks in $R$ are well separated, but their values are
much lower values than the BLG device.

These features in the resistance map of MLG are reflected in its
band structure, shown in Fig.~\ref{fig3}(b). As seen in the left
panel of Fig.~\ref{fig3}(b), bands with $k_x=0$ and with $k_x=1/2$
always cross $E_F$ at $k_y=0$ in MLG. Even though there is no gap
opening, placing the zero-energy level at this band crossing at
$k_y=0$ gives rise to a resistance maximum. In BLG, on the other
hand, the interlayer coupling opens local gaps around $k_y=0$,
resulting in a much higher resistance of the BLG in comparison to
the MLG device.

As seen in the middle panel of Fig.~\ref{fig3}(b), the MLG bands
at $k_y=0$ are highly dispersive and the energy spectrum is free
of gaps. Also, the states at both $(k_x=0,k_y=0)$ and
$(k_x=1/2,k_y=0)$ are doubly degenerate. Thus, $k_y=0$ states
contribute one conduction mode for all values of $V_{TG}$ and
$V_{BG}$. Absence of scattering in this periodically gated channel
is another demonstration of Klein tunneling in MLG.

We note that the double-degeneracy of these eigenstates of
free-standing MLG is protected in the 1D periodic potential
$\phi(x)$ by the symmetry operation
\begin{equation}
O = %
   \left(%
   \begin{array}{*{20}{c}} %
   1 & 0\\
   0 & -1
   \end{array}
   \right)%
   K \,,
\label{eq14}
\end{equation}
where $K$ is the complex conjugation operator. This can be
explained simply, since the Hamiltonian for MLG
\begin{equation}
 \tilde{H}_{MLG} = \left( %
 {\begin{array}{*{20}{c}} %
 \phi(x) & v_F{\bf{p}}^\dag \\
 v_F{\bf{p}} & \phi(x)
 \end{array}} %
 \right)
 \label{eq15}
\end{equation}
remains invariant under $O$, so that the degeneracy of the
above-mentioned eigenstates is not broken by any periodic
potential $\phi(x)$. We should also note that this symmetry
protection only occurs for electrons with $k_y=0$ corresponding to
normal incidence on the wires. There is no symmetry protection for
off-normal incidence, so that such electrons may be reflected,
giving rise to an interference pattern in the resistance map.
Nevertheless, since $k_y$ is near-zero for most electrons
contributing to transport in the device, most carriers are
transmitted and do not contribute to the interference pattern in
the resistance map. Since only a minority of electrons undergo
reflection and interference in MLG, corresponding resistance
maxima are less pronounced in the resistance map of MLG.

The situation is different in BLG, where the interlayer hopping
integral $\gamma_1$ breaks the $O$ symmetry. As seen in the middle
panel of Fig.~\ref{fig3}(a), BLG bands at $k_y=0$ show very little
dispersion along $k_x$ near $E_F$ due to the interlayer
interaction. Since these bands do not cross $E_F$, the
corresponding states do not contribute to conduction, thus
lowering the off-current and increasing the contrast in the
resistance map.

As mentioned earlier, BLG is doped by electrons at positive gate
voltages and by holes at negative voltages. Even though the
magnitude $|v_F|$ of the Fermi velocity does not depend on the
sign of the doping carriers, the direction of $\bf{v_F}$ in
electron-doped BLG is opposite to that of hole-doped BLG. In some
respect, this is parallel to the particle-hole symmetry found in
BLG and MLG.

\section{Summary and Conclusions}

In conclusion, we have studied the propagation of electrons in
periodically gated bilayer graphene as a way to construct a 2D
electronic metamaterial. We identified an intriguing
interference-like pattern, similar to that of a Fabry-P\'{e}rot
interferometer, in the resistance map in response to doping and
potential modulation provided by the extended bottom gate and the
periodic top gate. We provided a quantitative explanation for the
observations by considering quantum corrections to the
position-dependent potential in the channel region and the
equilibrium Green function formalism that describes ballistic
transport in BLG. We find periodically gated BLG to be a suitable
candidate for a distributed Bragg reflector for electrons.

\label{Acknowledgments}
\begin{acknowledgments}
This study has been inspired by Siqi Wang, Mervin Zhao, Changjian
Zhang, Sui Yang, Yuan Wang, Kenji Watanabe, Takashi Taniguchi,
James Hone, and Xiang Zhang, who kindly discussed their
experimental results with us. X.L. and D.T. thank Dan Liu for
fruitful discussions. X.L. acknowledges support by the China
Scholarship Council and the National Natural Science Foundation of
China (Grant No. 11974312). D.T. acknowledges financial support by
the NSF/AFOSR EFRI 2-DARE grant number EFMA-1433459. Computational
resources have been provided by the Michigan State University High
Performance Computing Center.
\end{acknowledgments}


%

\end{document}